\documentclass[prb,twocolumn,showpacs,aps,superscriptaddress,floatfix]{revtex4-2}
\usepackage{amsmath}
\usepackage{amssymb}
\usepackage{amsfonts}
\usepackage{bm}
\usepackage[dvips]{graphicx}
\usepackage{color}
\usepackage{ulem}
\usepackage{tabularx}
\usepackage[unicode, pdftex]{hyperref}
\usepackage{verbatim}
\DeclareGraphicsExtensions{.eps,.pdf}
\graphicspath{{pictures/}}
\usepackage{epstopdf}
\begin{document}
	\title{Pauli paramagnetism of triplet Cooper pairs in a nematic superconductor}
    \author{D. A. Khokhlov}
    \affiliation{Dukhov Research Institute of Automatics, Moscow, 127055 Russia}
    \affiliation{Moscow Institute of Physics and Technology, Dolgoprudny,
    Moscow Region, 141700 Russia}
    \affiliation{National Research University Higher School of Economics, 101000 Moscow, Russia}
   
    \author{R. S. Akzyanov}
    \affiliation{Dukhov Research Institute of Automatics, Moscow, 127055 Russia}
       \affiliation{Moscow Institute of Physics and Technology, Dolgoprudny,
    Moscow Region, 141700 Russia}
    \affiliation{Institute for Theoretical and Applied Electrodynamics, Russian
    Academy of Sciences, Moscow, 125412 Russia}

	\begin{abstract}
         We investigate the response of a doped topological insulator Bi$_2$Se$_3$ with spin-triplet nematic superconductivity to external magnetization.  We calculate the Zeeman part of magnetic susceptibility for nematic and chiral superconducting phases near T$_c$ in Ginzburg-Landau formalism. Superconducting order parameter from $E_u$ representation has non-trivial coupling with the transversal Zeeman field that results in a paramagnetic response to a magnetization. The topology of a Fermi surface has a strong influence on magnetic susceptibility. Lifshitz transition from closed to open Fermi surface eventually leads to phase transition from the nematic to chiral phase. At the transition point, magnetic susceptibility diverges. Also, we study the effects of the electron-electron interaction on the competition between nematic and chiral phases. We found that in a real system, electron-electron interaction can drive nematic to chiral phase only in the vicinity of the phase transition. We compare our results with the existing experimental data.  
    \end{abstract}

	\maketitle
\section{Introduction}
Several years ago topological superconductivity in the bulk of doped topological insulator (TI) A$_x$Bi$_2$Se$_3$ was     discovered~\cite{Hor2010,Sasaki2011,Kirzhner2012}. Dopant atom A can be Cu~\cite{Hor2010,Sasaki2011,Kirzhner2012,Kawai2020,Yonezawa2016,Tao2018,Matano2016}, Sr~\cite{Shruti2015,Liu2015,Kuntsevich2018,Kuntsevich2019,Pan2016,Neha2019} or Nb~\cite{Qiu2015,Kurter2018,Asaba2017,Das2020}. Superconducting phase in these materials shows C$_2$ rotational symmetry that breaks C$_3$ symmetry of the normal state. Such a rotational symmetry breaking was observed in measurements of specific heat~\cite{Yonezawa2016}, magnetic resonance~\cite{Asaba2017}, in-plane upper critical field~\cite{Pan2016,Kuntsevich2018}, vortex core form~\cite{Tao2018}. Observation of nuclear magnetic resonance indicates the spin-triplet character of this superconductivity~\cite{Matano2016}. 

Such C$_2$ rotational symmetry is possible due to the realization of the superconducting vector order parameter that belongs to E$_u$ representation of crystalline D$_{3d}$ point group~\cite{Fu2010, Fu2014}. One of possible solutions within such a representation is a nematic order parameter that is a two-component real vector $\pmb{\eta}=\eta(\cos\theta;\sin\theta)$. The direction of this vector is associated with the nematicity axis. Orientation of the nematicity axis shows the direction of two-fold symmetry breaking~\cite{Venderbos2016, Kuntsevich2019}. Nematic order parameter brings several interesting features such as vestigial nematic order~\cite{Hecker2018}, surface Andreev bound states~\cite{Hao2017}, unconventional Higgs mode~\cite{Uematsu2019}, and quasiparticle interference~\cite{Chen2018, Khokhlov2021}.     

Another possible solution for the order parameter is called chiral and corresponds to the vector with an imaginary component $\pmb{\eta}\propto (1;\pm i)$. This order parameter spontaneously breaks time-reversal symmetry, keeping rotational crystalline symmetry intact~\cite{Venderbos2016_2}. In Nb$_x$Bi$_2$Se$_3$ such spontaneous breaking of time-reversal symmetry was found~\cite{Qiu2015}. Muon spin rotational experiment shows time-reversal symmetry breaking in Sr$_{0.1}$Bi$_2$Se$_3$ superconductor~\cite{Neha2019}. However, another research of Nb$_x$Bi$_2$Se$_3$ shows the presence of time-reversal symmetry in the superconducting phase~\cite{Asaba2017, Kurter2018, Das2020}. Theoretical calculations predict the chiral phase is the ground state in 2D films of doped Bi$_2$Se$_3$~\cite{Venderbos2016_2, Chirolli2018} while in a 3D system with a closed Fermi surface, the nematic phase has the lower free energy than the chiral state~\cite{Chirolli2017, Yuan2017, Akzyanov2020}. Experimentally, the Fermi surface of doped Bi$_2$Se$_3$ with the superconductivity has a topology of an open cylinder~\cite{Almoalem2020, Lahoud2013} which is an intermediate case between 2D and 3D Fermi surface. This transition to the open Fermi surface leads to the phase transition from nematic to chiral phase~\cite{Yang2019}.  

Near the critical temperature, physical properties of the superconducting state are described by the
Ginzburg-Landau (GL) functional that for the nematic superconductor was obtained in Refs.~\cite{Fu2014, Venderbos2016}. One of the interesting features of this functional is the presence of the linear in magnetization's powers, coupling between the magnetic and superconducting degrees of freedom. Such coupling can lead to the transition from the nematic to chiral state with the spontaneous magnetization~\cite{Akzyanov2020_2}. It was predicted that transition from the nematic to chiral state occurs upon doping by magnetic ions~\cite{Chirolli2017, Yuan2017}. 

Magnetization measurements show the presence of the diamagnetic Meissner effect\cite{Das2020}. Muon spin rotational ($\mu$SR) experiment can be used to determine local magnetic moments in superconductors~\cite{Drew2009, Morenzoni2011, Khasanov2020, MacLaughlin1988}. 
Recent $\mu$SR experiment did not find time-reversal breaking in superconducting Nb$_{0.25}$Bi$_2$Se$_3$ in the absence of magnetic field~\cite{Das2020}. However, $\mu$SR shows that the superconducting state has an additional paramagnetic magnetization compared to the normal state in the magnetic field. In this work, we show how non-trivial coupling between magnetization and the nematic superconductivity can explain such a paramagnetic response.  

In this paper, we calculate the magnetic susceptibility of the nematic superconductor near the critical temperature $T_c$. We start with the microscopic derivation of the GL free energy of the doped topological insulator with the open Fermi surface and finite magnetization that is induced by the magnetic field following the common procedure~\cite{Venderbos2016, Chirolli2017, Yuan2017}. We solve GL equations for the order parameter and get that the magnetic field influences the form of the order parameter for the ground state. We calculate ground-state free energy as a function of a magnetic field. Using this free energy, we get that in both nematic and chiral states a paramagnetic contribution to the magnetic susceptibility exists in the system. We interpret this phenomenon as a Pauli paramagnetism of spin $1$ Cooper's pairs of the nematic superconductor. Such magnetic susceptibility diverges near the transition from the nematic to the chiral state. In addition, we show that coupling between magnetism and superconductivity is rather weak and electron-electron interaction can not cause the phase transition between nematic and chiral superconductivity.

The paper is organized as follows: in Sec.~\ref{suscp_model} we describe the normal and superconducting state in the presence of an external Zeeman field and calculate GL coefficients. Sec.~\ref{Sec::Pauli} is dedicated to the Zeeman susceptibility of the superconductor. In Sec.~\ref{suscp_coulomb} we consider the possibility of the phase transition due to electron-electron interaction. We discuss and summarize obtained results in Sec.~\ref{Sec::Discuss}.

\section{Model}\label{suscp_model}
\subsection{Normal phase}
We describe bulk electrons in a doped topological insulator of Bi$_2$Se$_3$ family by low-energy $\mathbf{k\cdot p}$ two-orbital Hamiltonian~\cite{Liu2010}: 
\begin{eqnarray}
    \hat{H}_{0}(\mathbf{k})=-\mu+m\sigma_z+v_zk_z\sigma_y+v(k_xs_y-k_ys_x)\sigma_x,
    \label{Eq::H0}
\end{eqnarray}
where $\mu$ is the chemical potential, $2m$ is a single-electron gap at zero chemical potential, Fermi velocities $v$ and $v_z$ describe motion in the $(\Gamma K;\Gamma M)$ plane and along $\Gamma Z$ direction correspondingly. Pauli matrices $s_i$ act in spin space while matrices $\sigma_i$ act in space of Bi and Se orbitals $\mathbf{p}=(P^1,P^2)$, where $i=\{x,y,z\}$, Planck constant $\hbar=1$. The Hamiltonian~(\ref{Eq::H0}) obeys time-reversal symmetry $\hat{\mathcal{T}} \hat{H}_0(\mathbf{k}) \hat{\mathcal{T}}^{-1}=\hat{H}_0(-\mathbf{k})$, where $\hat{\mathcal{T}}=is_y \hat{K}$, $\hat{\mathcal{T}}^2=-1$  is time-reversal operator and $\hat{K}$ provides complex conjugation. Also, this Hamiltonian has inversion symmetry $\hat{P}\hat{H}_0(\mathbf{k}) \hat{P}=\hat{H}_0(-\mathbf{k})$, where $\hat{P}=\sigma_z$, $\hat{P}^2=1$ is the operator of the inversion~\cite{Liu2010}. 

\begin{figure}[t]
    \center{\includegraphics[width=1\linewidth]{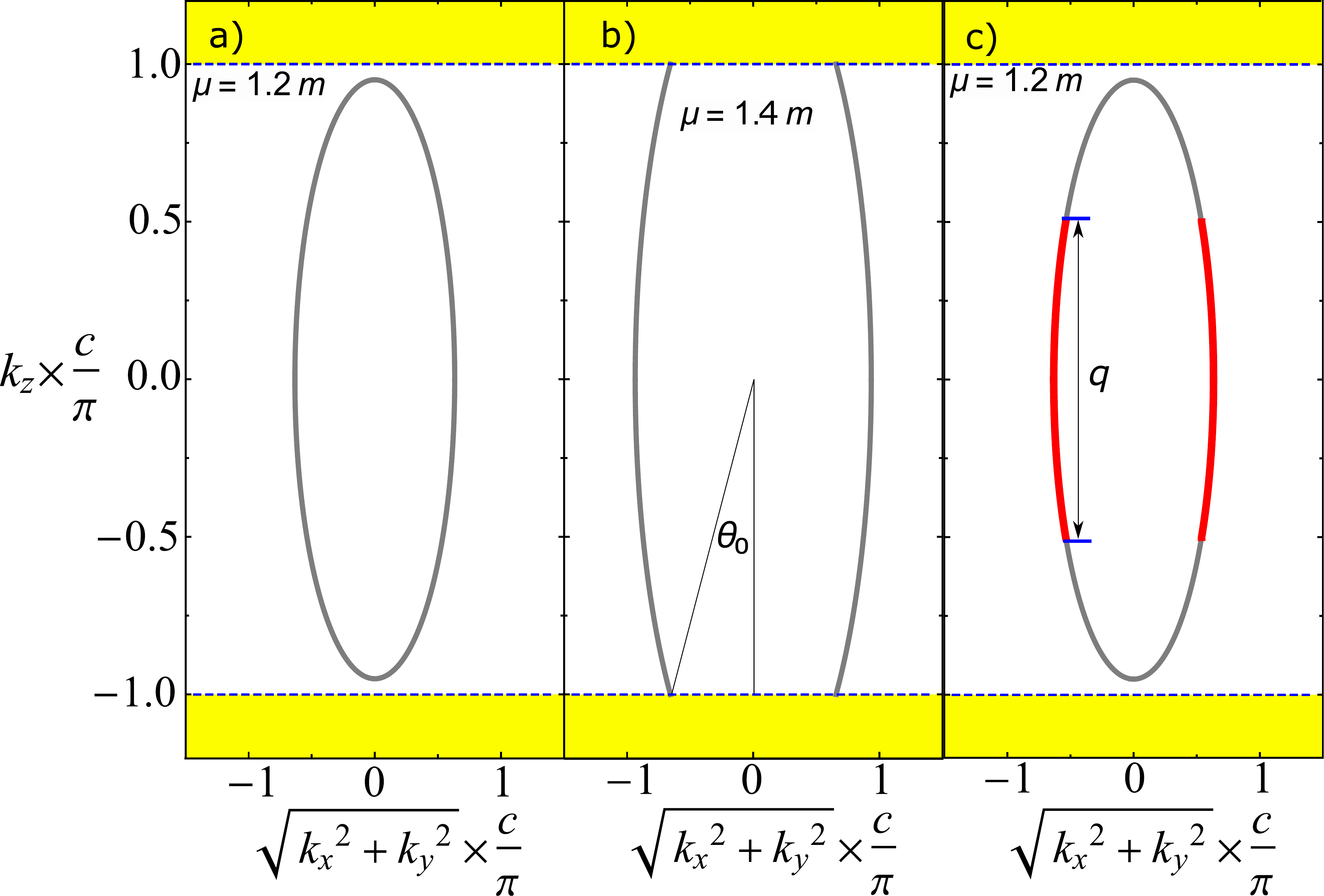}}
    \caption{Fermi surface before and after Lifshitz transition in dimensionless coordinates. The dashed blue line gives boundaries of the first BZ (white background) and second BZ (yellow background). Panel a): Gray curve gives closed Fermi surface at $\mu=1.2 m$. Panel b): Gray line gives open Fermi surface at $\mu=1.4 m$, cutoff angle $\theta_0$ is defined by Eq.~(\ref{Eq::cutoff_angle}). Panel c): The same Fermi surface as in panel a) where parts of Fermi surface with $|k_z|<q/2$ are highlighted by thick red lines. In these red sheets of FS, the electron-phonon coupling is the strongest, see Ref.~\cite{Wang2019} for details.}
    \label{Fig::FS}
\end{figure}
Angle-resolved photoemission spectroscopy and measurements of Shubnikov-de Haas oscillations show that the Fermi surface in doped Bi$_2$Se$_3$ is open in $\Gamma Z$-direction~\cite{Almoalem2020, Lahoud2013}. An increase of the chemical potential by doping transforms the closed ellipsoid Fermi surface of undoped samples to a corrugated cylinder Fermi surface of doped samples.
For high chemical potential, dispersion in $z$-direction disappears, and the system becomes effectively 2D. We introduce a finite length of 001 lattices constant $c$ into the model. Momentum is bounded by BZ size $|k_z|<\pi/c,$ or in elliptical coordinates $(vk_x,vk_y,v_z k_z)$ polar angle stays in range $\theta\in [\theta_0;\pi-\theta_0]$, where we define cutoff angle $\theta_0$ as follows   
\begin{eqnarray}
\cos(\theta_0)=\text{min}\left(1,\frac{\pi v_z}{c\sqrt{\mu^2-m^2}}\right).
\label{Eq::cutoff_angle}
\end{eqnarray}
In this model, the Fermi surface is closed at a low chemical potential $\mu$. When chemical potential reaches a critical value $\mu_{\text{LT}}=\sqrt{\pi^2v_z^2/c^2-m^2}$, the Fermi surface reaches the boundary of the first Brillouin zone (BZ) and changes topology from ellipsoid to open cylinder. That is the Lifshitz transition (LT)~\cite{Lifshitz1960}. We show closed and open Fermi surfaces in Figs.~\ref{Fig::FS}a) and b) correspondingly.  

Experiment with inelastic neutron scattering~\cite{Wang2019} shows linewidth of phonon spectra has singularity for phonon momentum $\mathbf{q}$ oriented along $\Gamma\text{Z}$ direction while $q\to 0$. The attraction between electrons with momenta $\mathbf{k}$ and $-\mathbf{k}$ in the Cooper channel occurs only for electrons whose $k_z$ components are close enough to each other. Thus, only electrons with the momentum $|k_z|<q/2$ participate in superconducting pairing. This singular coupling can be modeled by the Heaviside step function $\theta(q-2k_z)$ and the Fermi surface where Cooper pair's can be formed is effectively cut off. This new effective Fermi surface has the topology of an open cylinder, see Fig.~\ref{Fig::FS}. 
\subsection{Superconducting phase}\label{subsec::SC_phase}
We describe superconductivity in Nambu-II basis, where wave function is  
\begin{eqnarray}
    \Psi_{\mathbf{k}}=(\phi_{\mathbf{k}}^t,-i\phi_{\mathbf{k}}^{\dagger}s_y)^t,
\end{eqnarray}
where $\phi_{\mathbf{k}}=(\phi_{\uparrow,1,\mathbf{k}},\phi_{\downarrow,1,\mathbf{k}},\phi_{\uparrow,2,\mathbf{k}},\phi_{\downarrow,2,\mathbf{k}})^t$, symbol $t$ is transpose and symbol $^{\dagger}$ is Hermitian conjugate. Operator $\phi_{\uparrow(\downarrow),\sigma,\mathbf{k}}^{(\dagger)}$ annihilates (creates) electron with up (down) spin on the orbital $\sigma={P^1,P^2}$ with momentum $\mathbf{k}$. Superconducting order parameter from E$_u$ representation of D$_{3d}$ crystalline point group has the following matrix structure~\cite{Venderbos2016_2}:
\begin{eqnarray}
\label{Eq::Delta}   \hat{\Delta}=\eta_xs_x\sigma_y(\tau_x+i\tau_y)+\eta_x^*s_x\sigma_y( \tau_x-i\tau_y)+\\ \nonumber \eta_ys_y\sigma_y(\tau_x+i\tau_y)+\eta_y^*s_y\sigma_y(\tau_x-i\tau_y).
\end{eqnarray}
The superconducting term depends on two components of vector order parameters $\eta_x=\eta\sin(\alpha)e^{i\phi_1}$ and $\eta_y=\eta\cos(\alpha)e^{i\phi_2}$, where $\phi=\phi_1-\phi_2$. Matrices $\tau_i$ acts in electron-hole space. We assume that only the electrons in the Debye window participate in the superconductivity $-\omega_D<\epsilon_{\mathbf{k}}<\omega_D$, where $\epsilon_{\mathbf{k}}$ is the band's dispersion of the Hamiltonian~(\ref{Eq::H0}). This order parameter violate inversion symmetry of the normal state.

The BdG Hamiltonian in Nambu-II basis is~\cite{Fu2010} 
\begin{eqnarray}
\hat{H}_{BdG}(\mathbf{k})=\tau_z\hat{H}_0(\mathbf{k})+\hat{\Delta}.
\label{Eq::Hbdg}
\end{eqnarray}
We find GL free energy from microscopical theory as~\cite{Venderbos2016,Chirolli2017,Yuan2017} \begin{eqnarray}
\label{Eq::F_microscopic}
F=F_0- T\sum_{\omega}\int\frac{d\mathbf{k}^3}{(2\pi)^3}\text{Tr}\left[\log(1-\hat{G}_0\hat{\Sigma})\right],
\end{eqnarray}
where $F_0=-T \sum_{\omega}\int\frac{d\mathbf{k}^3}{(2\pi)^3}\text{Tr}\left[\log(\hat{G}_0^{-1})\right]$ is a free energy of a normal state, $\hat{\Sigma}$ is a self energy. Matrix  $\hat{G}_0=(i\omega-\hat{H}_{0})^{-1}$ is Matsubara Green's functions in the Nambu II basis and fermionic Matsubara frequency $\omega=(2n+1)\pi T$, where $n$ is integer. We take trace $\text{Tr}[..]$ over spin, orbital and electron-hole degrees of freedom. Then we expand logarithm from Eq.~(\ref{Eq::F_microscopic}) into Taylor series $\log(1-\hat{G}_0\hat{\Sigma})=-\sum_n \frac{(\hat{G}_0\hat{\Sigma})^n}{n}$  in powers of a perturbation $\hat{\Sigma}$ and combine terms of the series in powers of order parameters. Calculated GL coefficients are given in table~\ref{table:1}. Similar calculations are provided in Refs.~\cite{Yuan2017,Venderbos2016_2,Chirolli2017}. 

We start from $\hat{\Sigma}=\hat{\Delta}$ and obtain superconducting part of GL free energy~\cite{Venderbos2016_2} up to terms $\propto \eta^4$ 
\begin{eqnarray}
F_{sc}(\eta_x,\eta_y)=A(|\eta_x|^2+|\eta_y|^2)\!+\!B_1(|\eta_x|^2+|\eta_y|^2)^2\!+\\ \nonumber
B_2|\eta_x^*\eta_y-\eta_x\eta_y^*|^2. 
\label{Eq::GL_F}
\end{eqnarray}
The coefficient $A\propto(T-T_c)$ changes sign from positive to negative under cooling from a temperature above $T_c$ to a temperature under $T_c$ inducing superconducting phase transition~\cite{Venderbos2016_2}. 

In the chiral phase expression $|\eta_x^*\eta_y-\eta_x\eta_y^*|=\eta^2$, while in the nematic phase this expression is zero. Thus, the free energy of the chiral phase has an additional term $B_2\eta^4$. Therefore, in the system with positive $B_2>0$, the nematic phase is the ground state, while negative $B_2<0$ promotes the chiral phase. The sum of coefficients $B_1+B_2$ is always positive. Direct calculations in model with infinite Brillouin zone and the Hamiltonian~(\ref{Eq::H0}) show $B_2>0$ in 3D system~\cite{Chirolli2017} and $B_2<0$ in 2D case~\cite{Chirolli2018}. 
\begin{figure}[t]
    \center{\includegraphics[width=1\linewidth]{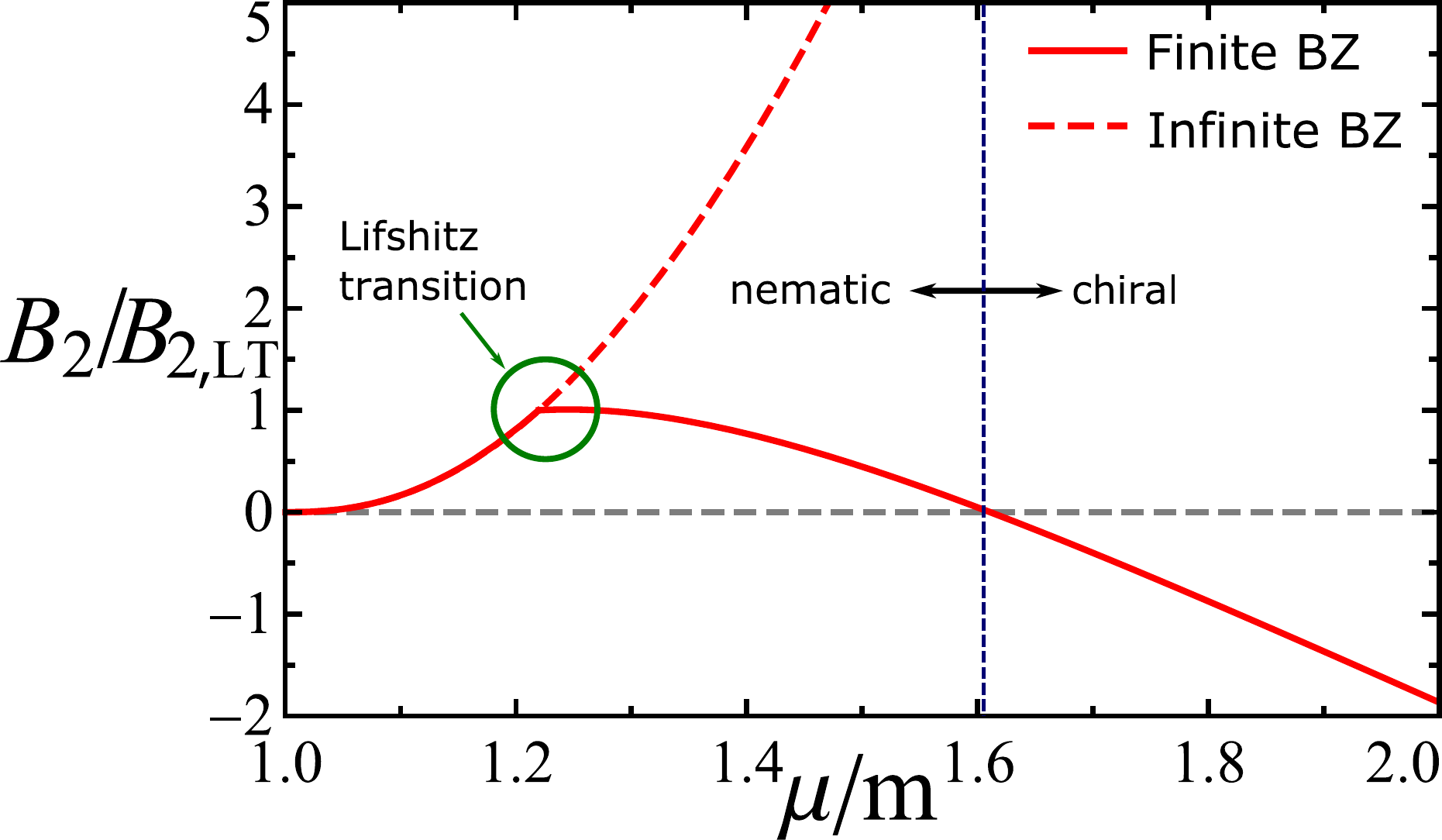}}
    \caption{GL coefficient $B_2(\mu)$ in units of $B_{2,LT}$ at the Lifshitz transition point vs dimensionless chemical potential $\mu/m$. The solid red curve corresponds to $B_{2}$ calculated in a model with finite BZ. Dash red curve corresponds to $B_{2}$ calculated in model with infinite BZ. Dash gray horizontal line corresponds to $B_2=0$. The vertical dash blue line shows the transition between the nematic phase with $B_2>0$ and the chiral phase with $B_2<0$.}
    \label{Fig::B2}
\end{figure}

In the absence of magnetism, nematic phase has a two-component vector order parameter with real components $\pmb{\eta}=\eta(\cos\alpha;\sin\alpha)$. In chiral phase order parameter is a two-component vector again with one real component and another one pure imaginary $\pmb{\eta}=\frac{\eta}{\sqrt{2}}(1;\pm i)$, which breaks time-reversal symmetry. 

We calculate coefficient $B_2$ for arbitrary lattice constant $c$ in a model with finite BZ. Results are given in Table~\ref{table:1}. The coefficient $B_2$ is plotted as a function of Fermi energy $\mu$ in Fig.~\ref{Fig::B2} in models with finite and infinite BZ. Fermi surface is closed at low chemical potential even when BZ is finite and both models are the same. In the model with an infinite BZ, coefficient $B_2>0$ grows with an increase in Fermi energy. Thus nematic order parameter has lower energy than the chiral one. Lifshitz transition occurs at $\mu_{LT}=\sqrt{\frac{\pi^2 v_z^2}{c^2}+m^2}$ in model with finite BZ and leads to decline in $B_{2}$. This coefficient reaches zero in point $\mu^*=\sqrt{\frac{v_z^2\pi^2}{c^2\cos^2(\theta_0^*)}+m^2}$, where angle $\theta_0^*\approx0.985$ does not depend on any parameters. Critical chemical potential $\mu^*$ indicates phase transition between nematic and chiral superconductivity. In all numerical calculations we set superconducting critical temperature $T_c=10^{-3}m$, in-plane velocity $v=1/3m c$ and $\Gamma Z$-velocity $v_z=\frac{2}{3}v$. 
\subsection{Coupling between the magnetism and superconductivity}
In presence of transversal magnetic field electrons in doped Bi$_2$Se$_3$ experience orbital depended Zeeman magnetization. We decompose this magnetization into ferromagnetic (FM) and antiferromagnetic (AFM) with corresponding Lande g-factors $\beta_f$ and $\beta_a$, see Ref.~\cite{Liu2010} and write it down as
\begin{eqnarray}
\hat{\Sigma}_{m}=\mu_B\beta_f H s_z+\mu_B\beta_a H s_z\sigma_z.
\label{Eq::magnetic_pert}
\end{eqnarray}

\begin{figure}[t]
    \center{\includegraphics[width=1\linewidth]{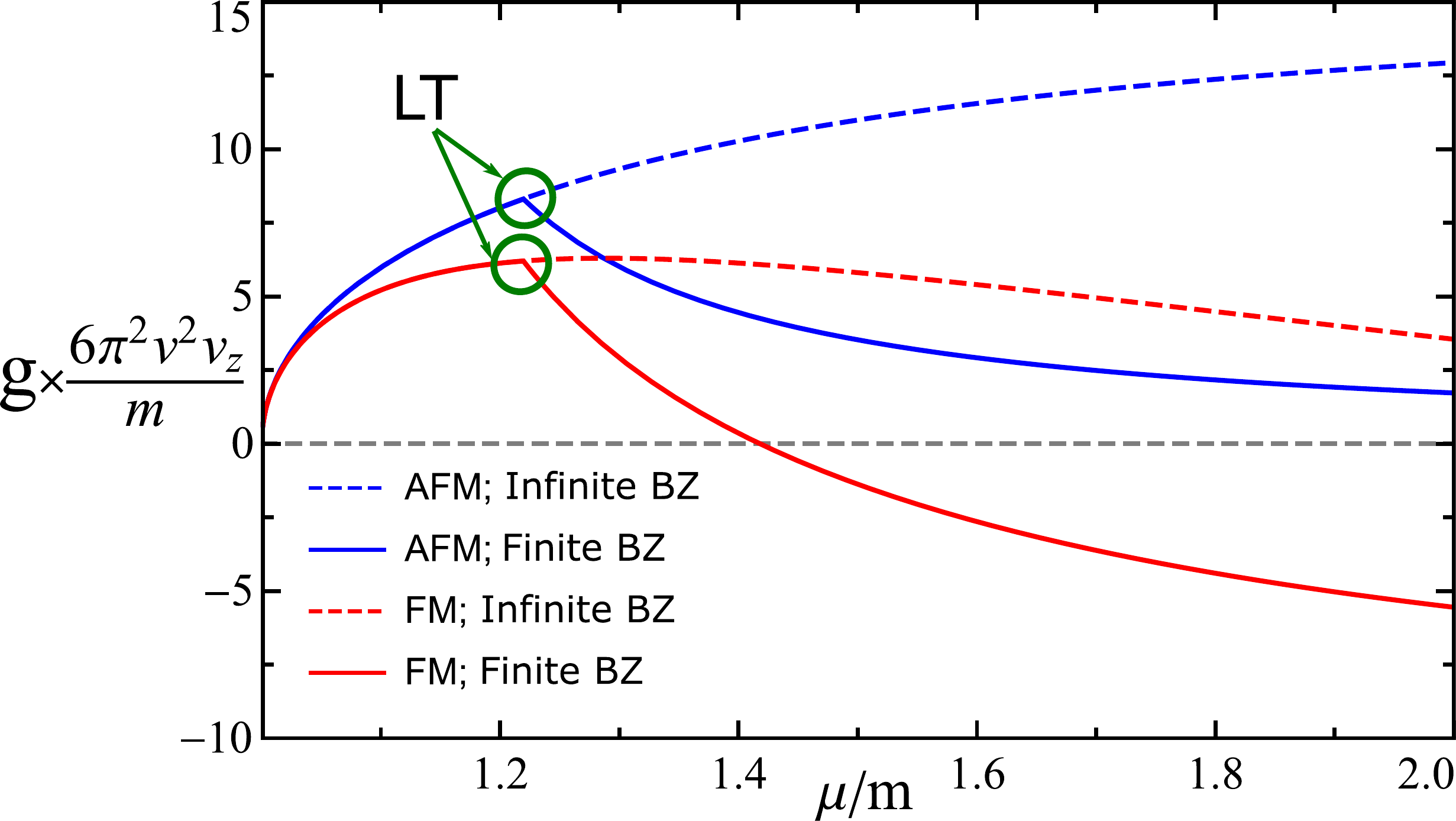}}
    \caption{GL coefficients g$_f(\mu)$ and g$_a(\mu)$ in units of $\frac{m}{6\pi^2v^2v_z}$ versus dimensionless chemical potential $\mu/m$. Red curves correspond to ferromagnetic coupling g$_f$ while blue curves corresponds to antiferromagnetic coupling $g_a$. Solid lines correspond to calculations in a model with finite BZ. Dash curves correspond to calculations in a model with infinite BZ. Dash gray line indicates zero level. The green circles mark Lifshitz transition.}
    \label{Fig::g}
\end{figure}

We calculate free energy via Eq.~(\ref{Eq::F_microscopic}), where perturbation is $\hat{\Sigma}=\hat{\Sigma}_m+\hat{\Delta}$. We expand $\log(1-\hat{G}_0\hat{\Sigma})$ in powers of $H$ and $\eta$ up to $O(\hat{\Sigma}^2)$. The free energy receive additional terms~\cite{Yuan2017,Chirolli2017}
\begin{eqnarray}
\label{Eq::FM_free_energy}
F_{\alpha}(H,\eta_x,\eta_y)=-2ig_{\alpha}\mu_B\beta_{\alpha}H(\eta_x^*\eta_y-\eta_x\eta_y^*)+\\ \nonumber a_{\alpha}(\mu_B\beta_{\alpha}H)^2,
\end{eqnarray}
\begin{eqnarray}
\label{Eq::FA_coupling} 
&F_{m}(H)=a_{m}\beta_f\beta_a\mu_B^2H^2,
\end{eqnarray}
where $\alpha=\{f,a\}$ refers to FM or AFM terms. Constants $a_f, a_a$ and $a_m$ describes response of the normal state to a transverse magnetization, see Table~\ref{table:1}. We combine terms that $\propto H^2$ into one effective term $a_{eff}\mu_B^2H^2=(a_{f}\beta_f^2+a_{a}\beta_a^2+a_{m}\beta_f\beta_a)\mu_B^2H^2$. 

Coefficients $g_f$ and $g_a$ describe the coupling between magnetism and superconductivity. Note, that in-plane magnetization does not produce such terms. We calculate these coefficients in models with finite and infinite BZ, see Fig~\ref{Fig::g}. Ferromagnetic coefficient $g_f$ changes sign with the increase of the chemical potential, while antiferromagnetic coefficient $g_a$ is always positive. According to expression for $g_f$ at arbitrary geometry of Fermi surface given in Table~\ref{table:1}, it changes sign $g_f=0$ if $\cos^2(\theta_0^{**})=\frac{3\left(\log(j\omega/T) \left(m^2-\mu^2\right)+m^2\right)}{-3\mu^2\log(j\omega/T)+2\mu^2+m^2}$. In this point coupling between superconductivity and FM perturbation disappears. 
\begin{figure}[t]
    \center{\includegraphics[width=1\linewidth]{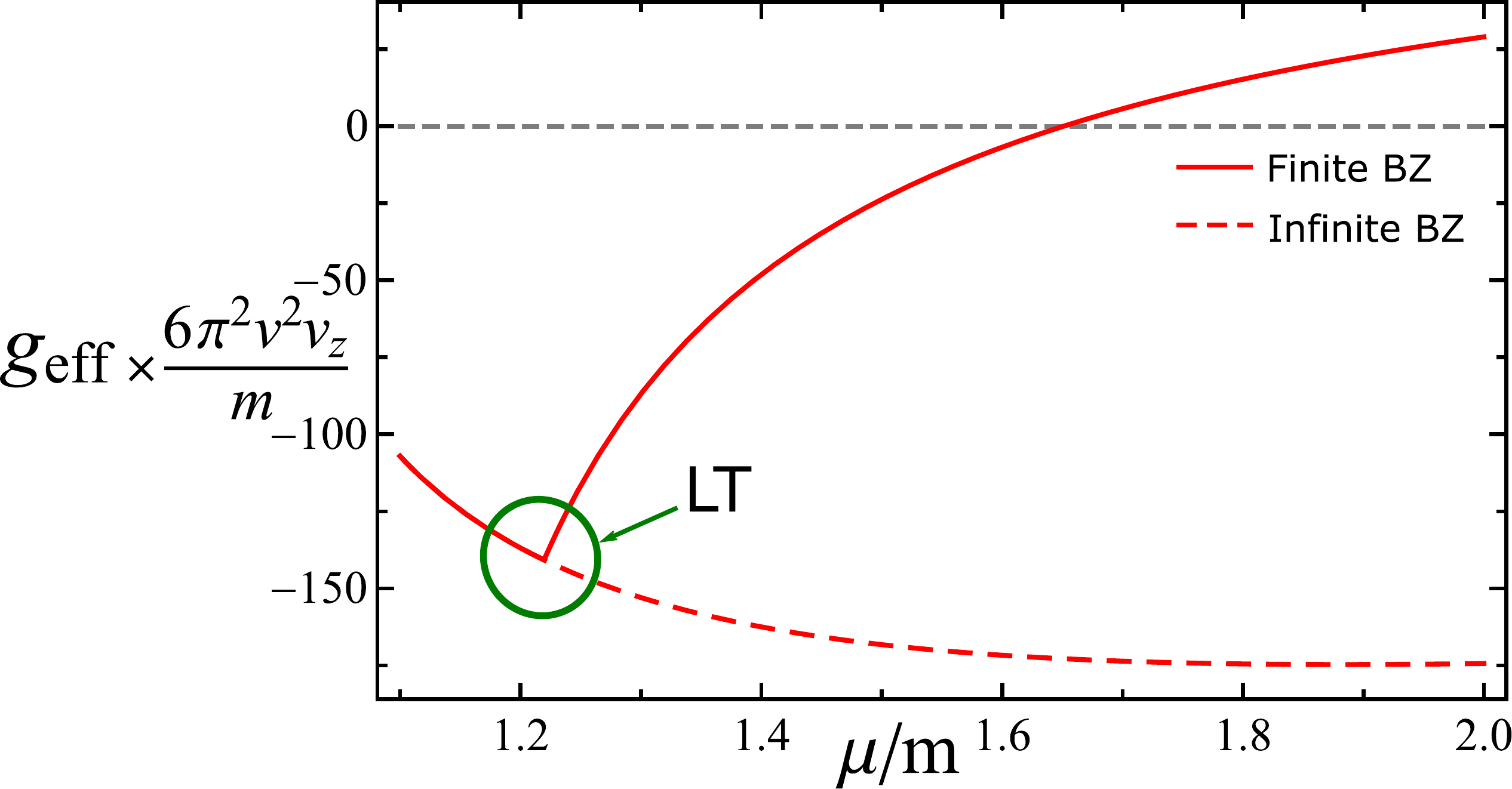}}
    \caption{Effective coupling between superconductivity and Zeeman field g$_{eff}(\mu)$ in units of $\frac{m}{6\pi^2v^2v_z}$ versus dimensionless chemical potential $\mu/m$. The solid red curve gives g$_{eff}$ calculated in a model with finite BZ, while dash red curves correspond to calculations in a model with infinite BZ. Dash gray curve indicates zero level. The green circle marks Lifshitz's transition.}
    \label{Fig::g_eff}
\end{figure}
The total coupling between Zeeman field and superconductivity is described by effective coefficient $g_{eff}=\beta_f g_f+\beta_a g_a$. We plot $g_{eff}(\mu)$ in Fig.~\ref{Fig::g_eff}. Coefficient $g_{eff}$ is negative at low chemical potential since both FM and AFM g-factors are negative. Then $g_{eff}$ change sign from negative to positive at some value of Fermi energy. Zero-point of $g_{eff}$ occurs at higher chemical potential than zero-point of $g_f$ in model with open FS. In numerical calculations we set Lande g-factors $\beta_f=-5.3$, $\beta_a=-7.4$.
\begin{table*}[h!]
\small
\begin{tabular}{|m {1.3cm}|m {4.4cm}| m {4.4cm}| m {8.2cm}|} 
 \hline
 GL coeff. & $\theta_0=0$: Closed FS & $\theta_0=\pi/2$: cylindrical FS & $0<\theta_0<\pi/2$: Open FS \\
 \hline
 $$A$$ & $$\frac{\sqrt{\mu^2-m^2}^3(T-T_c)}{3\pi^2\mu T_c v^2v_z}$$ & $$\frac{(\mu^2-m^2)(T-T_c)}{4\pi c\mu T_c v^2}$$ & $$\frac{\sqrt{\mu^2-m^2}^3(T-T_c)\cos(\theta_0)(7+\cos(2\theta_0))}{24\pi^2\mu T_c v^2v_z}$$\\ 
 \hline
 $$B_1$$ & $$\frac{7\zeta(3)\sqrt{\mu^2-m^2}^5}{30\pi^4T_c^2\mu^3v^2v_z}$$ & $$\frac{21\zeta(3)(\mu^2-m^2)}{128\pi c T_c^2\mu^3v^2}$$ & $$\frac{7\zeta(3)\cos(\theta_0)\sqrt{\mu^2-m^2}^5(427+76\cos(2\theta_0)+9\cos(4\theta_0))}{15360\pi^2T_c^2\mu^3v^2v_z}$$ \\
 \hline
 $$B_2$$ & $$\frac{7\zeta(3)\sqrt{\mu^2-m^2}^5}{60\pi^4T_c^2\mu^3v^2v_z}$$ & $$-\frac{7\zeta(3)(\mu^2-m^2)}{128\pi c T_c^2\mu^3v^2}$$ & $$\frac{7\zeta(3)\cos(\theta_0)\sqrt{\mu^2-m^2}^5(71+188\cos(2\theta_0)-3\cos(4\theta_0))}{15360\pi^2T_c^2\mu^3v^2v_z}$$ \\
 \hline
 $$g_f$$ & $$\frac{\sqrt{\mu^2-m^2}}{6\pi^2\mu^2v^2v_z}\times$$ $$\Bigg(3m^2\log\left(\frac{j\omega_D}{T_c}\right)+2(m^2-\mu^2)\Bigg)$$ &
 $$\frac{1}{2\pi c v^2\mu^2}\times$$
 \begin{eqnarray*}
    &\Bigg(m^2+(m^2-\mu^2)\log\left(\frac{j\omega_D}{T_c}\right)\Bigg)
 \end{eqnarray*}
 &
 $$\frac{\sqrt{\mu^2-m^2}\cos(\theta_0)}{6\pi^2v^2v_z\mu^2}\times$$
 \begin{eqnarray*}
    &\Bigg\{m^2\left[3-\cos^2(\theta_0)+3\log\left(\frac{j\omega_D}{T_c}\right)\right]-\\
    &\mu^2\Bigg[2\cos^2\theta_0+ 3\sin^2\theta_0\log\left(\frac{j\omega_D}{T_c}\right)\Bigg]\Bigg\}
 \end{eqnarray*}
 \\
 \hline
 $$g_a $$ & $$\frac{m\sqrt{\mu^2-m^2}}{6\pi^2\mu^3v^2v_z}\times$$ $$\Bigg(3\mu^2\log(j\omega_D/T_c)+\mu^2-m^2\Bigg)$$ &  
 $$\frac{m\sqrt{\mu^2-m^2}}{2\pi c v^2\mu}$$
 & 
 $$-\frac{m\sqrt{\mu^2-m^2}\cos(\theta_0)}{6\pi^2v^2v_z\mu^3}\times$$
 \begin{eqnarray*}
    &\Bigg(m^2\cos^2(\theta_0)-\mu^2\Bigg[2-\cos(2\theta_0)+3\cos^2\theta_0\log\left(\frac{j\omega_D}{T_c}\right)\Bigg]\Bigg)
 \end{eqnarray*}
 \\
 \hline
 $$a_{f,z} $$ &  
  $$-\frac{\sqrt{\mu^2-m^2}(2m^2+\mu^2)}{12\pi^2v^2v_z\mu}$$ & $$-\frac{m^2}{4\pi v^2c\mu}$$ 
 & 
 $$-\frac{\sqrt{\mu^2-m^2}\cos(\theta_0)(5m^2+\mu^2+\cos(2\theta_0)(\mu^2-m^2))}{24\pi^2v^2 v_z \mu}$$
 \\
 \hline
 $$a_{a,z} $$ &  
  $$-\frac{\sqrt{\mu^2-m^2}(m^2+2\mu^2)}{12\pi^2v^2v_z\mu}$$ & $$-\frac{\mu}{4\pi v^2c}$$ 
 & 
 $$-\frac{\sqrt{\mu^2-m^2}\cos(\theta_0)(m^2+5\mu^2-\cos(2\theta_0)(\mu^2-m^2))}{24\pi^2v^2 v_z \mu}$$
 \\
 \hline
 $$a_{a,f} $$ & \multicolumn{3}{c|}{
    \begin{minipage}{\linewidth}
     $$\frac{m}{\pi^2\mu}$$
    \end{minipage}}\\ \hline
    \multicolumn{4}{|c|}{$a_{\alpha,e}=1/V_{\alpha}+a_{\alpha,z}$} \\ \hline
\end{tabular}
\caption{GL coefficients $A$, $B_1$, $B_2$, $g$ and $a$ for systems with different Fermi surface and different magnetic interaction. Left column describes 3D system with closed FS. Coefficients in 2D system are shown in central column. Right column gives general case of open FS. Coefficient $j=\frac{2}{\pi}e^c$, where $c$ is Euler–Mascheroni constant and $\zeta$ is Riemann zeta function.}
\label{table:1}
\normalsize
\end{table*}

\section{Magnetic susceptibility} \label{Sec::Pauli}
Response of the superconductor to a magnetic field can be decomposed into a response to orbital and Zeeman parts of the magnetic field. Orbital part induces Meissner currents and dominates overall response. Zeeman part occurs due to change of the order parameter in case of finite magnetization. In this paper, we focus only the second term. Total GL free energy of $E_u$ superconductor in the Zeeman field is
\begin{eqnarray}
\label{Eq::F_eta_H}
\!\!\!\!\!\!\!\!\!\!\!F(H,\eta_x,\eta_y)\!=\!F_{sc}(\eta_x,\eta_y)\!- \!2ig_{eff}\mu_B H(\eta_x^*\eta_y\!-\!\eta_x\eta_y^*)\!+\!\\  \nonumber
a_{eff} H^2.
\end{eqnarray}
The free energy depends on the value of complex values components of the order parameter $\eta_x$ and $\eta_y$ as well as it depends on external magnetic field $H$. We minimize free energy~(\ref{Eq::F_eta_H}) as a function of $\pmb{\eta}$ and find equilibrium free energy in given Zeeman field $H$. The exact expression for the equilibrium free energy depends on the sign of the $B_2$ coefficient. We start from nematic case $B_2>0$. Finite $H$ breaks time-reversal symmetry and for the ground state $\sin(2\alpha)\sin(\phi)\propto H$ becomes non zero, however we still name this phase nematic~\cite{Akzyanov2020_2}. In this case free energy is written as
\begin{eqnarray}
\label{Eq::F_H}
\!\!\!\!\!\!\!\!F_{min}^{nem}(H)\!=\!\!-\frac{A^2}{4B_1}\!+\!a_{eff}H^2\!\!-\!\frac{g_{eff}^2H^2}{B_2}\!\!\left(\!\!1\!\!-\!\!\frac{g_{eff}^2}{B_2a_{eff}}\!\!\right)\!\!.
\end{eqnarray}
The first term gives free energy of the nematic phase in a zero field. The second term describes diamagnetism of the normal state. The third term appears due to coupling between the nematic superconductivity and Zeeman field and describes the superconducting Pauli paramagnetism.  We set $\left(1-\frac{g_{eff}^2}{B_2a_{eff}}\right)\sim\left(1- \frac{T_c^2}{\mu^2}\right)\simeq 1$ since coupling $g_{eff}$ is small enough. This approximation is valid everywhere besides the small neighborhood of point $B_2=0$.

Zeeman part of magnetic susceptibility is 
\begin{eqnarray}
\chi_{ns}=-\frac{\partial^2F_{min}^{nem}(H)}{\partial H^2}=-2a_{eff}+\frac{2g_{eff}^2}{B_2}.
\end{eqnarray}
Here the first term corresponds to Zeeman susceptibility of normal phase, and the second term arises due to coupling between nematic superconductivity and magnetism. Zeeman susceptibility has a jump under phase transition between normal and nematic superconducting phases
\begin{eqnarray}
\chi_{ns}-\chi_n=\frac{g_{eff}^2\mu_B^2}{2B_2}>0.
\end{eqnarray}
This jump is relatively small compared to the Pauli paramagnetism of the normal phase $\chi_n$ with a small parameter $(\chi_{ns}-\chi_n)/\chi_n=g_{eff}^2/(4B_2a_{eff})\sim (T_c/\mu)^2$. However, this additional paramagnetism is enhanced close to area $B_2=0$, where phase transition between nematic and chiral phases occurs. 

Now we minimize free energy~(\ref{Eq::F_eta_H}) as a function of $\eta_x$ and $\eta_y$ for $B_2<0$ In this case order parameter chiral and free energy is  
\begin{align}
\label{Eq::F_H_ch}
\!\!\!\!\!\!\!\!F_{min}^{ch}(H)\!=\!\!\frac{-A^2}{\!4(B_1\!+\!B_2)}\!+\!\frac{Ag_{eff}H}{B_1\!+\!B_2}\!-\!\frac{g_{eff}^2H^2}{B_1+B_2}+\!a_{eff}H^2\!.\!\!
\end{align}
Jump of susceptibility between the chiral superconducting and normal phase is 
\begin{eqnarray}
    \chi_{cs}-\chi_n=\frac{g_{eff}^2}{4(B_1+B_2)}>0,
\end{eqnarray}
which has no singularities in contrast with $\chi_{ns}$ since $B_1+B_2>0$.   
\begin{figure}[t]
    \center{\includegraphics[width=1\linewidth]{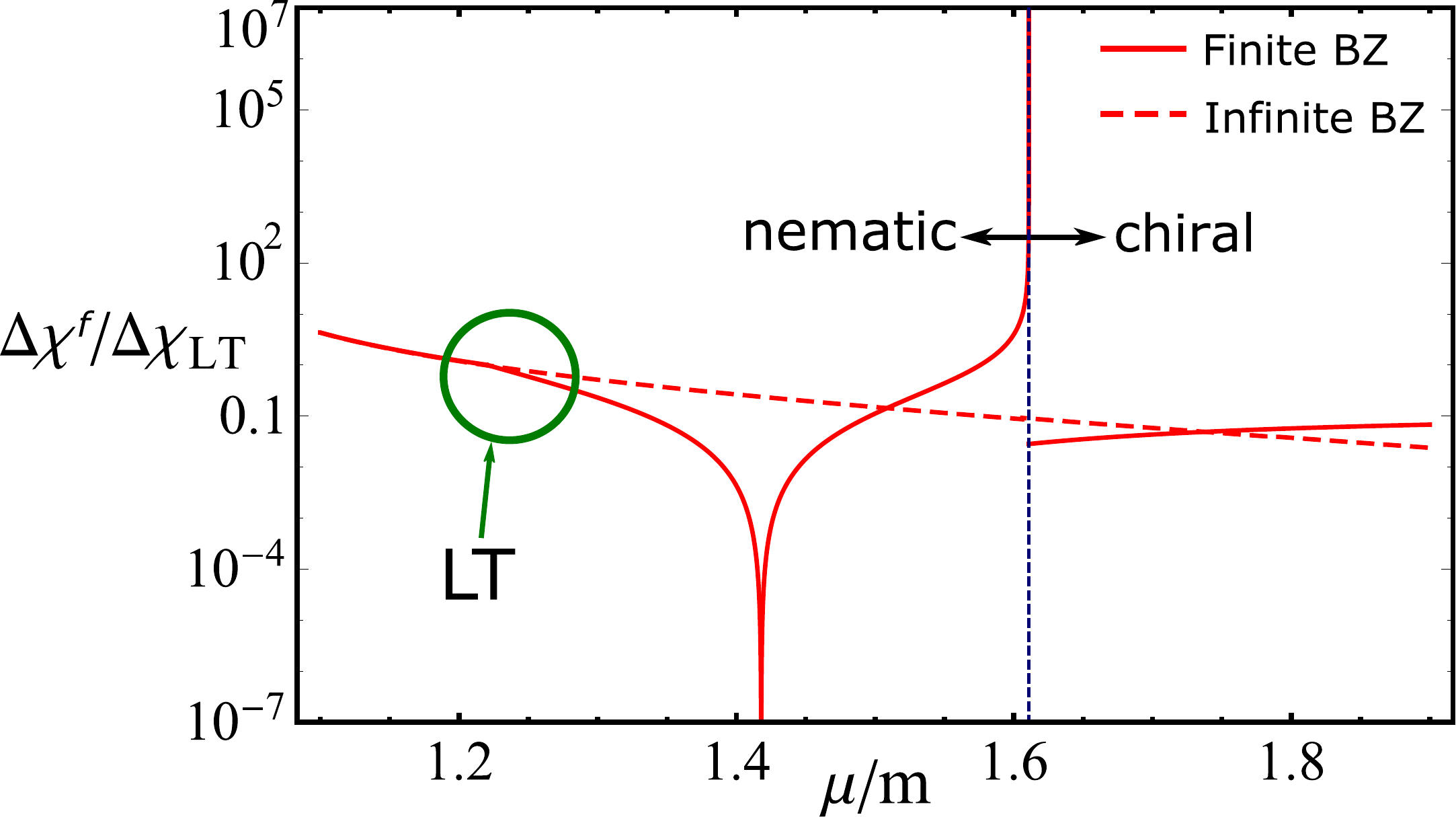}}
    \caption{Jump of Zeeman susceptibility between superconducting  and normal phases $\Delta\chi^{f}$ normalized by its value $\Delta\chi^{f}_{LT}$ in Lifshitz transition point as a function of the dimensionless chemical potential $\mu/m$ in model with only one Lande g-factor $\beta_f=-5.3$. The red solid curve corresponds to finite BZ. The red dashed curve corresponds to infinite BZ. Green circle indicates Lifshitz transition. The vertical blue dashed line shows the transition between nematic and chiral phases.}
\label{Fig::Delta_chi_F}
\end{figure}
\begin{figure}[t]
    \center{\includegraphics[width=1\linewidth]{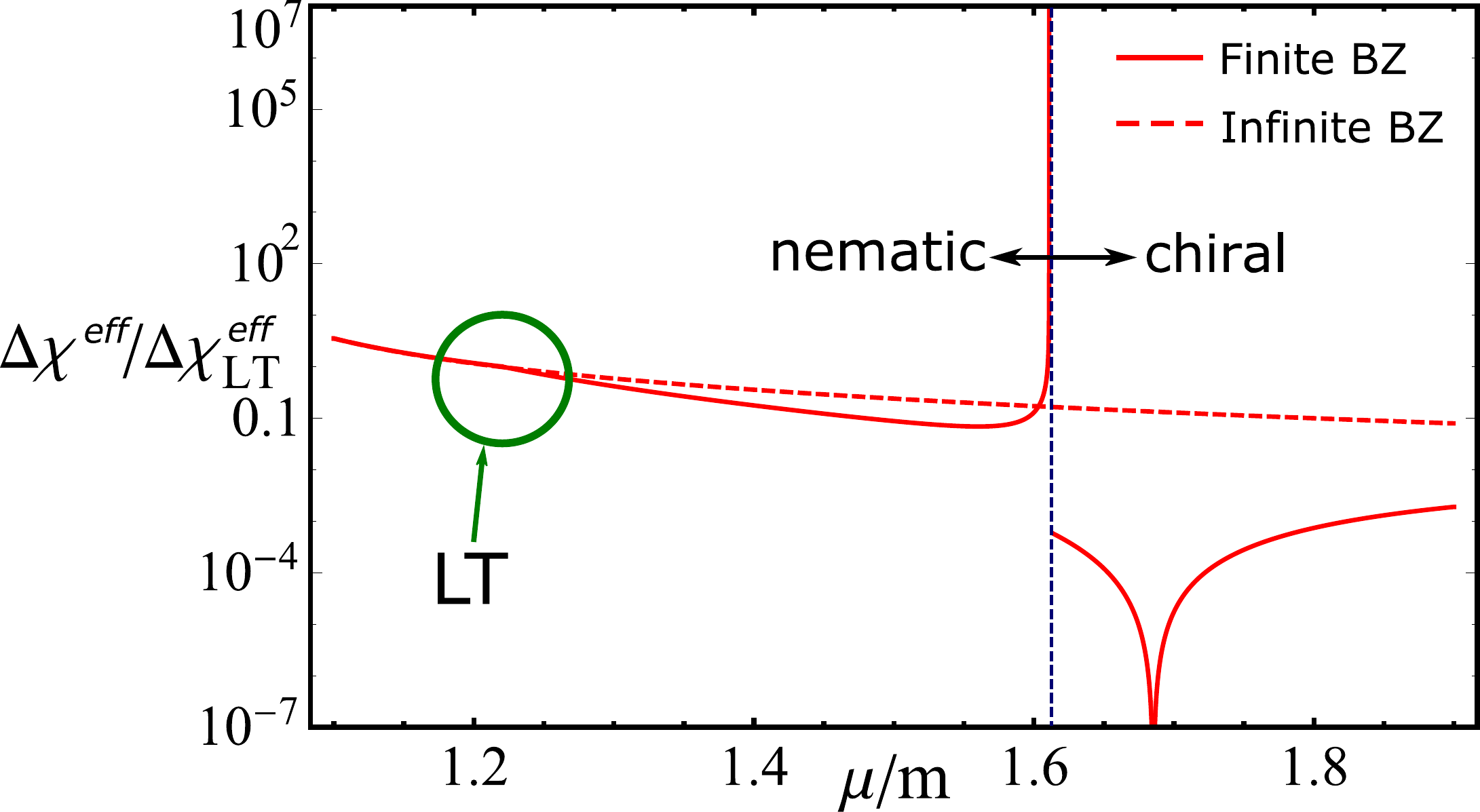}}
    \caption{Jump of Zeeman susceptibility between superconducting and normal phases $\Delta\chi^{eff}$ in units of its value $\Delta\chi^{eff}_{LT}$ in Lifshitz transition point as a function of dimensionless chemical potential $\mu/m$ in model with both Lande g-factors $\beta_f=-5.3$ and $\beta_a=-7.4$. The red solid curve corresponds to finite BZ. The Red dashed curve corresponds to infinite BZ. Green circle indicates Lifshitz transition. The vertical blue dashed line shows the transition between nematic and chiral phases.}
\label{Fig::Delta_chi_F_A}
\end{figure}

We calculate the Zeeman susceptibility as a function of Fermi energy. We consider two different cases when Zeeman field appears (i) only in FM channel with $g_f=-5.3$ and $g_a=0$, see Fig.~\ref{Fig::Delta_chi_F} and (ii) in both FM and AFM channels with $g_f=-5.3$ and $g_a=-7.4$, see Fig.~\ref{Fig::Delta_chi_F_A}. We see that the inclusion of AFM response to the magnetic field significantly changes the magnetic susceptibility.

In the model with finite BZ, Lifshitz transition occurs at $\mu_{LT}\approx1.25$. Both susceptibilities $\Delta\chi^{f}$ and $\Delta\chi^{eff}$ diverges for $\mu\approx 1.6$, where $B_2$ changes sign and phase transition from nematic to chiral phase occurs. Susceptibility $\Delta\chi^f$  vanishes in nematic phase in point $\mu\approx 1.4$, while $\Delta\chi^{eff}$ vanishes in chiral phase in point $\mu\approx 1.8$. This disappearance of susceptibilities occurs in points, where $g_f$ and $g_{eff}$ reaches zero value, see Figs.~\ref{Fig::Delta_chi_F},~\ref{Fig::Delta_chi_F_A}. The model with an infinite BZ system always stays in the nematic phase. Thus no divergence in susceptibility occurs.

Superconductor with E$_u$ pairing has triplet structure i.e. Cooper pairs have spin equal to one~\cite{Fu2014}. It can be seen if we write down order parameter using second quantization form $\eta_x \propto c_{1\uparrow}c_{2\uparrow}+c_{1\downarrow}c_{2\downarrow}$ and $\eta_y \propto i(c_{1\uparrow}c_{2\uparrow}-c_{1\downarrow}c_{2\downarrow})$ where $c_{is}$ is the operator of the annihilation of the electron on $i$-th orbital with the spin projection $s$.  We can see that the order parameter represents two Cooper's pairs with opposite spins $\pm 1$. Each Cooper's pair couples electrons with the same spin on different orbitals. In presence of the magnetic field number of spin up and spin down Cooper's pairs is not equal. Such spin imbalance influences the form of the order parameter that defines the response to the magnetic field. 

Spin density of the system is expressed through full Green's function as $S_z=-T\sum_{\omega}\int\frac{d^3\mathbf{}k}{(2\pi)^3} \text{Tr}[\hat{G} s_z]$. We expand superconducting Green's function $\hat{G}\approx \hat{G}_0+\hat{G}_0\hat{\Delta}\hat{G}_0+\hat{G}_0\hat{\Delta} \hat{G}_0\hat{\Delta}\hat{G}_0$ in powers of superconducting order parameter~\ref{Eq::Delta} that gives us $S_z=-2ig_{eff}(\eta_x^*\eta_y-\eta_x\eta_y^*)\propto \sin(2\alpha)\sin(\phi)\propto n_{\uparrow 1}n_{\uparrow 2}-n_{\downarrow 1}n_{\downarrow 2}$ where $n_{s \sigma}=\phi^{\dagger}_{s\sigma}\phi_{s\sigma}$ is density of electrons from orbital $\sigma$ with spin $s$~\cite{Fu2014}. We see that parameter $\sin(2\alpha)\sin(\phi)$ is responsible for spin imbalance of Cooper pair's. While the order parameter is real, angle $\phi=0$, spin density is zero and number of spin up and spin down Cooper's pairs is equal. Turning on Zeeman field makes product $\sin(2\alpha)\sin(\phi)\sim H$ and spin density that arises due to imbalance between spin up and spin down Cooper's pairs becomes $S_z=\frac{g_{eff}^2 \mu_B H}{2B_2}\sim k_F^3\beta_{eff}^2(\mu_B H/\mu)(T_c/\mu)^2$. Such spin imbalance decreases the energy of the system in a magnetic field similarly to the Pauli paramagnetism of the electrons. Thus, we refer this effect to a Pauli paramagnetism of Cooper pairs.

Average spin density of Cooper pairs in chiral phase is $S_z=g_{eff}\eta \approx Ag_{eff}/(B_1+B_2)\sim\delta T\cdot T_c k_F^3/\mu^2$. Therefore, chiral phase has nonzero magnetization even in absence of magnetic field. For comparison, spin density of (i) a normal-phase Bi$_2$Se$_3$ and (ii) a fully spin-polarized metal in Zeeman field $H$ have order of (i)  $k_F^3\beta_{eff}\mu_BH/\mu$ and (ii) $k_F^3$ correspondingly. 
Density of Cooper pairs in chiral phase enlarges in presence of Zeeman field $\eta=\frac{A+2|g_{eff}H|}{2(B_1+B_2)}$. Thus, spin density receives a term $\frac{g_{eff}^2\mu_B H}{4(B_1+B_2)}$ and a sample receives extra polarization proportional to Zeeman field $H$. Therefore, chiral phase has Zeeman susceptibility despite been fully spin-polarized.

Note that the chiral phase degenerates regarding the sign of the spin density. The magnetic field removes such degeneracy, and the chiral phase with the spins of Cooper's pairs aligned with the magnetic field becomes a ground state. If the sign of the spin density is fixed by some other field, then the response to the magnetic field that aligned against the spins of Cooper's pairs will be diamagnetic.
\section{Phase transition under electron-electron repulsion}\label{suscp_coulomb} 
Coupling between magnetic and superconducting order parameters can lead to the emergence of ferromagnetic or antiferromagnetic phase in the superconducting state. We consider effects of point-like electron-electron repulsion. We treat this interaction in the mean-field approach. Only FM and AFM order parameters couples with $E_u$ superconductivity $M_{z}=\langle n_{\uparrow}\rangle-\langle n_{\downarrow}\rangle$ and $L_{z}=\langle n_{\uparrow,1}\rangle+\langle n_{\downarrow,2}\rangle-\langle n_{\uparrow,2}\rangle-\langle n_{\downarrow,1}\rangle$ correspondingly. Here $n_{s\sigma}$ is local density of electrons with spin $s$ from orbital $\sigma$. Both order parameters enter in free energy in the same way as Zeeman field~(\ref{Eq::FM_free_energy}) where factors $\mu_B H \beta_{f(a)}$ should be replaced by $V_{f(a)}M_z(L_z)$. Coefficients $g_{f(a)}$ and $a_m$ do not change while coefficient $a_{f(a)}\to a_{f(a),e}=a_{f(a)}+1/V_{f(a)}$, see table~\ref{table:1}. This coupling may inspire the phase transition from nematic to chiral phase that can occur even under $B_2>0$. Particularly while $g_a=0$ phase transition occurs when $\frac{g_f^2}{a_{f,e}B_2}>1$, see Ref.~\cite{Akzyanov2020_2}. Direct calculations show the coefficient order $(T/\mu)^2\ll1$. While both FM and AFM order parameters exist, we find a generalized condition on the phase transition
\begin{eqnarray}
\lambda=\frac{a_{a,e}g_f^2-a_m g_f g_a+a_{f,e}g_a^2}{(4a_{f,e}a_{a,e}-a_m^2)B_2}>1.
\label{Eq::phase_transition_condition}
\end{eqnarray}
Since AFM coupling $g_a\sim g_f$, generalized condition is met only in neighbourhood of a point $B_2=0$ or point $4a_{f,e}a_{a,e}-a_m^2=0$. The condition~(\ref{Eq::phase_transition_condition}) means the system stays extremely close to the appearance of magnetic order parameters even in the normal phase that seems unrealistic for a real system. Such a degree of proximity to the phase transition imposes strong conditions on electron-electron interaction $V_f$ and $V_a$. Therefore, the emergence of the chiral phase from the nematic phase under electron-electron interaction in the absence of magnetic ordering in the normal phase seems unlikely. For example, we find $\lambda\approx 6\cdot10^{-7}\ll 1$ when coupling constants are $V_f=200\text{eV}\cdot\text{\AA}$ and $V_a=100\text{eV}\cdot\text{\AA}$, Fermi energy $\mu/m=1.3$ and other parameters the same as at the end of Sec.~\ref{subsec::SC_phase}. Thus, electron-electron interaction cannot drive the system from the nematic to the chiral state for the experimentally achievable conditions.

\section{Discussion}\label{Sec::Discuss}

Recently, superconducting powder of Nb$_{0.25}$Bi$_2$Se$_3$ was investigated in $\mu$SR experiment~\cite{Das2020}. The authors measured the magnetization of a sample at different temperatures in the external magnetic field. In contrast with previous experiment~\cite{Qiu2015}, here authors find the system has time-reversal symmetry in the absence of a magnetic field. In Fig.3b,Ref.~\cite{Das2020} authors show temperature dependence of magnetization in superconducting phase counted from magnetization in normal phase at $T=5K$ (above $T_c$). Surprisingly, the magnetic moment of the superconducting phase is higher than in the normal one, which indicates the appearance of an additional paramagnetism associated with superconductivity. We claim that Pauli paramagnetism of triplet Cooper's pairs is observed in this experiment. Note that measurements of the magnetic susceptibility of the large monocrystal Nb$_{0.25}$Bi$_2$Se$_3$ using SQUID show a strong diamagnetic response that is expected in superconductors~\cite{Das2020}. 

Typically, superconductors in the low magnetic fields demonstrate diamagnetism due to Meissner effects, which occurs due to response to an orbital part of a magnetic field~\cite{Akzyanov2021, Schmidt2020}. In this paper, we focus on the coupling between spin and magnetization and do not take into account Meissner currents since they are calculated in previous works. Due to rather weak coupling between magnetization and superconductivity, we expect that diamagnetic susceptibility from the Meissner effect will dominate paramagnetic susceptibility for large samples. However, the Meissner effect can be suppressed for large thin films and near the critical temperature $T_c$.

Magnetism can induce phase transition between nematic and chiral superconductivity~\cite{Chirolli2017, Yuan2017}. We investigate this phase transition, assuming magnetism appears because of electron-electron repulsion with order parameters in FM and AFM channels. We find general condition on the phase transition~(\ref{Eq::phase_transition_condition}). Since coupling between magnetism and superconductivity is small, this condition is not met, and the phase transition can not be induced by electron-electron repulsion. 

In Ref.~\cite{Lahoud2013} evolution of the Fermi surface of Bi$_2$Se$_3$ upon doping was investigated. Authors showed that doping increases carrier density that leads to the transformation of closed Fermi surface to open. This transformation coincides with the emergence of the superconductivity and occurs at carrier density $2 \cdot10^{19}\text{cm}^{-3}\lesssim n\lesssim 10^{20}\text{cm}^{-3}$. According to Ref.~\cite{Almoalem2020} Lifshitz transition occurs at carrier density is $n=5\cdot 10^{19}\text{cm}^{-3}$. In Ref.~\cite{Kawai2020} authors investigated excessive Cu doping in Bi$_2$Se$_3$. They obtain several samples with Cu concentration $0.28<x<0.54$. According to Knight shift measurements, carrier density at $x>0.37$ strongly increases, which perhaps turns the system to the chiral phase. Thus, the realization of the chiral superconductivity in overdoped samples is possible in such samples. Since doping is non-uniform in general, macroscopic parts of the superconductor can be in the vicinity of the nematic to chiral phase transitions. Since paramagnetic susceptibility diverges near the transition, such parts can give a substantial contribution to the susceptibility that can be used to track phase transition.

\section*{Acknowledgment}
Authors acknowledge support by the Russian Science Foundation under Grant No 20-72-00030 and partial support from the Foundation for the Advancement of Theoretical Physics and Mathematics “BASIS”. 

\bibliography{nematic_PM}
\end{document}